\documentclass[aps,pra,superscriptaddress,reprint,floatfix,longbibliography,showpacs]{revtex4-1}
\usepackage[english]{babel}
\usepackage{amsmath}
\usepackage{amssymb}
\usepackage{graphicx}
\usepackage{hyperref}
\usepackage{upgreek}
\usepackage{epstopdf}
\usepackage{color}
\usepackage{ulem}
\usepackage{dcolumn}
\usepackage{multirow}
\usepackage[note-name=, use-sort-key = false]{notes2bib}

\def\cm{cm$^{-1}$}

\begin{document}
\title{Optical study of the charge dynamics evolution in the topological insulators MnBi$_2$Te$_4$ and Mn(Bi$_{0.74}$Sb$_{0.26}$)$_2$Te$_4$ under high pressure}


\author{M. K\"opf}
\affiliation{Experimentalphysik II, Institute of Physics, Augsburg University, 86159 Augsburg, Germany}
\author{S. H. Lee}
\affiliation{2D Crystal Consortium, Materials Research Institute, Pennsylvania State University, University Park, PA 16802, USA}
\affiliation{Department of Physics, Pennsylvania State University, University Park, Pennsylvania 16802, USA}
\author{Z. Q. Mao}
\affiliation{2D Crystal Consortium, Materials Research Institute, Pennsylvania State University, University Park, PA 16802, USA}
\affiliation{Department of Physics, Pennsylvania State University, University Park, Pennsylvania 16802, USA}
\affiliation{Department of Materials Science and Engineering, Pennsylvania State University, University Park, Pennsylvania 16802, USA}
\author{C. A. Kuntscher}\email{christine.kuntscher@physik.uni-augsburg.de}
\affiliation{Experimentalphysik II, Institute of Physics, Augsburg University, 86159 Augsburg, Germany}

\begin{abstract}
The van der Waals material MnBi$_2$Te$_4$ and the related Sb-substituted compounds Mn(Bi$_{1-x}$Sb$_x$)$_2$Te$_4$ are prominent members of the family of magnetic topological insulators, in which rare quantum mechanical states can be realized. In this work, we study the evolution of the charge dynamics in MnBi$_2$Te$_4$ and the Sb-substituted compound Mn(Bi$_{1-x}$Sb$_x$)$_2$Te$_4$ with $x=0.26$ under hydrostatic pressure.
For MnBi$_2$Te$_4$, the pressure dependence of the screened plasma frequency, the dc conductivity, and the reflectance at selected frequencies
shows weak anomalies at $\sim$2 and $\sim$4~GPa, which might be related to an electronic phase transition driven by the enhanced interlayer interaction.
We observe a pressure-induced decrease in the optical gap, consistent with the decrease in and closing of the energy gap reported in the literature.
Both studied materials show an unusual decrease in the low-energy optical conductivity under pressure, which we attribute to a decreasing spectral weight of the Drude terms describing the free charge carrier excitations. Our results suggest a localization of conduction electrons under pressure, possibly due to hybridization effects.
\end{abstract}

\maketitle


\section{Introduction}
In recent years, the famous representative MnBi$_2$Te$_4$ (MBT) of topological insulators with magnetic order was studied extensively with many different experimental techniques. It is proven by now that MBT is a promising candidate for hosting rare quantum mechanical effects such as
the quantum anomalous Hall effect (QAHE) or the axion insulator, which makes MBT potentially interesting for applications in quantum metrology and spintronics~\cite{Zhang.2019, Deng.2020, Lei.2021, He.2020, Li.2019, Li.2019a}. Characteristic for a topological insulator, MBT exhibits a gapped electronic structure near the Fermi energy $E_{\mathrm{F}}$ in the bulk, while linear dispersing, gapless surface states are induced by band inversion, which are protected by spatial and time-reversal symmetry at ambient conditions~\cite{Hao.2019, Chen.2019, Chen.2019a, Otrokov.2019, Klimovskikh.2020, Swatek.2020, Vidal.2019}. Whether the surface states are gapped or not, has been a matter of controversy, as contradictory experimental results were reported \cite{Lee.2019, Otrokov.2019a}.
Due to the onset of an $A$-type antiferromagnetic ordering below $T_{\mathrm{N}}=25\,$K~\cite{Li.2020, Chen.2019a, Ding.2020, Yuan.2020, Rani.2019, Cui.2019}, a time-reversally broken system in coexistence with nontrivial band topology emerges in MBT, which enables the realization of Weyl fermions close to $E_{\mathrm{F}}$~\cite{Lee.2021, Li.2020, Li.2019b, Li.2020a, Tokura.2019}.

\begin{figure*}[t]
	\includegraphics[width=\linewidth]{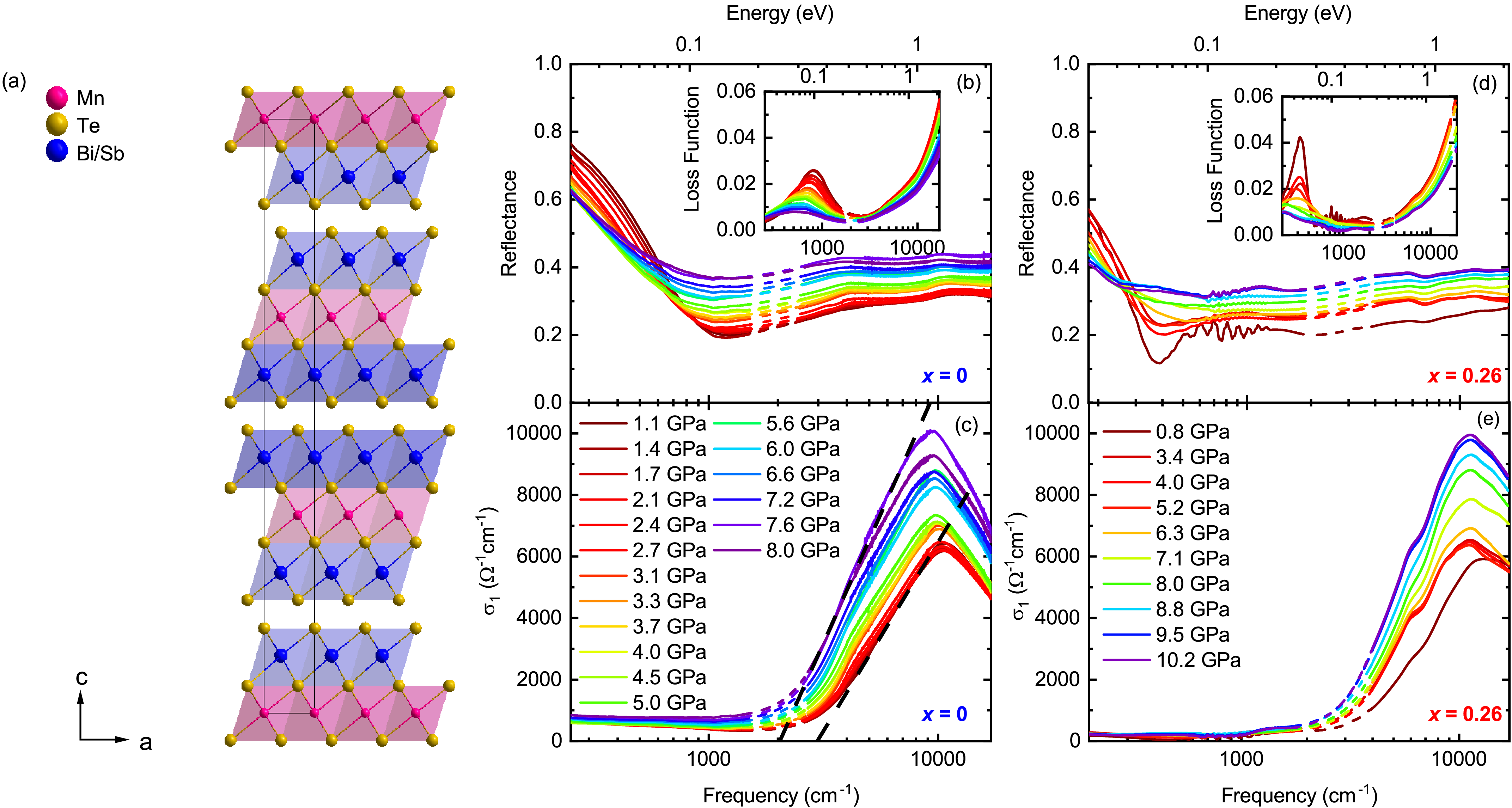}
	\caption{\label{fig.opticaldata} (a) Sketch of the crystal structure of MBST. Pressure-dependent (b) reflectance and (c) optical conductivity $\sigma_1$ of the pure compound as well as (d) reflectance and (e) optical conductivity $\sigma_1$ of the Sb-substituted compound with $x=0.26$. The corresponding loss functions of both materials are shown as an inset in (b) and (d), respectively.
In (c), the intercept point of the linear extrapolations (dashed lines) of $\sigma_1$ with the frequency axis indicates the onsets of interband transitions, corresponding to the optical gap size.}
\end{figure*}

\begin{figure*}[t]
	\includegraphics[width=1\linewidth]{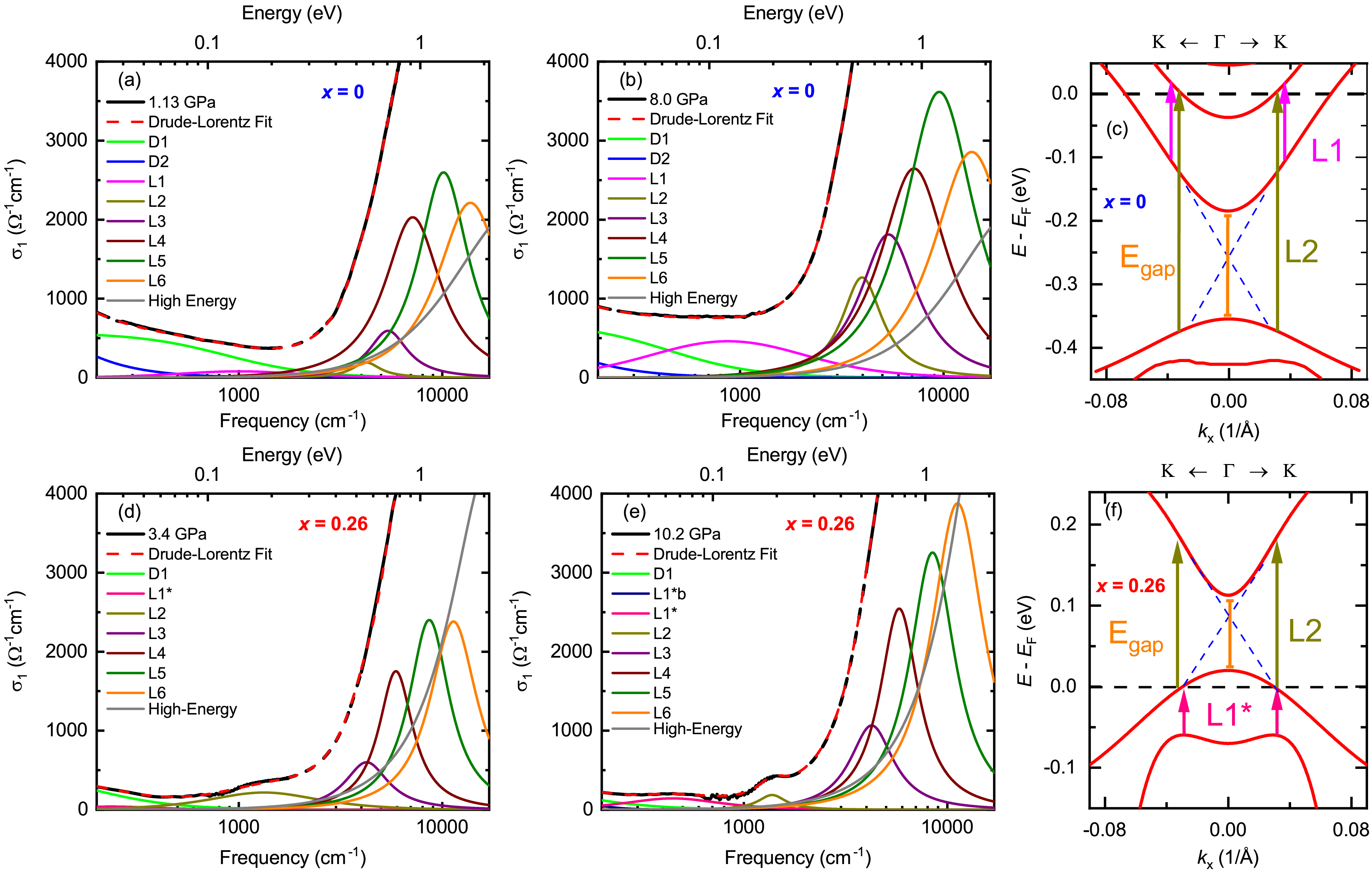}
	\caption{\label{fig.fitting} Drude-Lorentz fit of $\sigma_1$ of the pure compound at (a) 1.13\,GPa and (b) 8.0\,GPa as well as of the 26\% substituted compound at (d) 3.4\,GPa and (e) 10.2\,GPa. (c) and (f) display sketches of the corresponding band structures close to the Fermi level, respectively (similar to Ref.~\cite{Koepf.2022a} and based on calculations by Chen et al.~\cite{Chen.2019a}). In (c) and (f), the vertical arrows display electronic transitions described by the respective Lorentz oscillators, where L1 describes transitions between two conduction bands, L1* between valence and conduction band, and L2 corresponds to transitions across the optical gap. The energy gap $E_{\mathrm{gap}}$ is indicated by a vertical orange line.}
\end{figure*}

MBT crystallizes in the space group $R\overline{3}m$ with the cell parameters $a=4.33\,$\AA\ and $c=40.91\,$\AA~\cite{Li.2020} and it possesses a van-der-Waals type layered structure, consisting of Te-Bi-Te-Mn-Te-Bi-Te septuples stacked in a rhombohedral ABC order~\cite{Li.2020, Hou.2020, Rani.2019}. A sketch of the structure is shown in Fig.\,\ref{fig.opticaldata}\,(a). While the Mn sublattice contributes a magnetic moment to the compound, the Bi and Te $p_z$ bands are responsible for the nontrivial surface states, which are located at the $\Gamma$-point in momentum space~\cite{Lee.2019, Lai.2021}. Regarding the three-dimensional properties, a bulk energy gap between 180 and 220\,meV is reported~\cite{Chen.2019, Chen.2019a}, while the surface Dirac point between the valence and conduction band is located app. 270\,meV below the $E_{\mathrm{F}}$~\cite{Chen.2019a, Hao.2019, Swatek.2020}. A sketch of the described band structure is shown in Fig.\,\ref{fig.fitting}\,(c), where two electronic bands are crossing $E_{\mathrm{F}}$. Accordingly, MBT is intrinsically doped, with electrons as the main charge carriers \cite{Lee.2021,Yan.2019a}.

For further understanding of achieving ideal Weyl semimetal states, ways of tuning the electronic structure in MBT have been exploited~\cite{Lee.2021, Li.2021}. By gradually exchanging Bi atoms by Sb atoms, the Sb-substituted compounds Mn(Bi$_{1-x}$Sb$_x$)$_2$Te$_4$ (MBST) are generated, where the position of the chemical potential as well as the energy gap size are strongly affected and determined by the Sb content~\cite{Chen.2019, Lee.2021, Ko.2020, Yan.2019a, Riberolles.2021}. While the structural parameters undergo only minor changes due to Sb doping, a lowering of the $E_{\mathrm{F}}$ level from the conduction bands towards the valence bands with increasing Sb content can be observed, where a nearly insulating state is expected to be reached at a doping level of $x=0.26$~\cite{Chen.2019, Lee.2021}. Accordingly, the expected electronic band structure of Mn(Bi$_{0.74}$Sb$_{0.26}$)$_2$Te$_4$ ($x=0.26$) is sketched in Fig.\,\ref{fig.fitting}\,(f), where $E_{\mathrm{F}}$ now slightly crosses one of the lower bands, as will be justified later based on our experimental results. In addition, the energy gap is reported to be reduced with Sb doping level, where a gap closing is expected at $x=0.55$, and a reopening of the bands results in the vanishing of non-trivial bands for Sb-substituted compounds above $x=0.55$~\cite{Chen.2019}. Yet, also topological surface states have been detected in MnSb$_2$Te$_4$ ($x=1$) other studies~\cite{Wimmer.2021, Liu.2021, Eremeev.2021, Murakami.2019}.

Thus far, the optical properties of pure MBT and of various Sb-substituted compounds MBST have been investigated at ambient pressure only~\cite{Xu.2021, Koepf.2020, Koepf.2022, Koepf.2022a}. As unusual properties of pressurized MBT regarding dc electric transport dynamics, electronic structure, and structual evolution have been reported~\cite{Pei.2020, Guo.2021, Chen.2019b, Xu.2022, Shao.2021, Chong.2023}, it is important to also investigate the frequency-dependent optical response of MBT at high hydrostatic pressures. Typically, materials show a decreasing resistivity under external pressure, as bond lengths and interatomic distances are more and more compressed~\cite{Hazen.2009}. This holds for example for the closely related compounds and topological insulators Bi$_2$Te$_3$~\cite{Matsubayashi.2014} and MnSb$_2$Te$_4$~\cite{Yin.2021}. In stark contrast, several studies reveal a more complex behaviour of the dc resistivity for MBT under hydrostatic pressure~\cite{Pei.2020, Chen.2019b}. Electrical resistivity measurements show the initial {\it increase} in resistivity with increasing pressure up to $\sim$ 12\,GPa, followed by a strong resistivity drop for higher pressures. This unconventional, non-monotonic shift was suggested to result from the competing nature of gradually localized surface electrons and the bulk electrons undergoing a delocalization process under pressure~\cite{Pei.2020}.
In another study, the hybridization of Bi-6$p$ and Te-5$p$ electrons with delocalized Mn-3$d$ electrons, which creates a hybridization gap,
or an increase in electron scattering rate were given as possible mechanisms for the initial pressure-induced resistivity increase~\cite{Chen.2019b}.
Regarding the crystal structure under pressure, the cell parameters $a$ and $c$ shrink irregularly with increasing pressure~\cite{Chen.2019b}. This results in an unusual evolution in the $c/a$ ratio, where an initial drop to a minimum value at 3\,GPa is followed by an increase for higher pressures~\cite{Pei.2020}.
A further crucial consequence of external pressure application is a decrease in the energy gap at the $\Gamma$-point up to $\sim$ 15\,GPa~\cite{Pei.2020, Xu.2022}, as demonstrated by various theoretical calculations~\cite{Pei.2020, Xu.2022, Guo.2021}. In summary, the high-pressure properties of MBT are highly unusual and many open questions remain.

In this study, we investigate the effect of external pressure on the charge carrier dynamics in pure MBT and in the Sb-substituted compound Mn(Bi$_{0.74}$Sb$_{0.26}$)$_2$Te$_4$ by optical spectroscopy over a broad frequency range.
According to our optical data, the metallic character of both compounds weakens under pressure.
We also find anomalies in the pressure dependence of the screened plasma frequency $\omega_{\mathrm{pl}}^{\mathrm{scr}}$, dc conductivity, and reflectance values at selected frequencies. These pressure-induced anomalies are in good agreement with the structual parameter evolutions under pressure reported in the literature~\cite{Chen.2019b, Pei.2020}. Furthermore, we observe a decrease in the optical gap, which we relate to the decreasing energy gap under pressure~\cite{Pei.2020, Xu.2022}.

\section{Materials and Methods}
Single crystals of MnBi$_2$Te$_4$ and MnBi($_{0.74}$Sb$_{0.26}$)$_2$Te$_4$ were grown by the self-flux method as reported in Ref.~\cite{Lee.2021}. The 26\,\% Sb substituted sample belongs to the batch SL3B1, as described in our previous publication~\cite{Koepf.2022a}, and accordingly is slightly hole doped. The crystals have been characterized in detail by temperature-dependent electron transport and Hall resistivity measurements, as reported in Ref.~\cite{Lee.2021}. We carried out infrared reflectance measurements by Fourier-transform infrared (FTIR) spectroscopy with a Bruker 80v spectrometer coupled to a Hyperion microscope.

Hydrostatic pressure has been applied with EasyLab diamond anvil cells (DAC), where a small piece of the single crystal was placed into a hole of a CuBe gasket together with some CsI, which served as pressure transmitting medium for the realization of quasi-hydrostatic pressure. For the pure compound we have used a DAC with a 900\,$\upmu$m culet size and for the 26\,\% substituted sample a DAC with a 800\,$\upmu$m culet size has been selected. In order to determine the pressure inside the DAC, tiny Al$_2$O$_3$ spheres were put next to the sample. Excited with a green laser, these spheres emit radiation with a specific wavelength characteristic for the applied pressure. This radiation is detected with a charge-coupled device (CCD) spectrograph.

The reflectance is calculated according to the equation $I_{\mathrm{sample}}/I_{\mathrm{reference}}$, where $I_{\mathrm{sample}}$ stands for the intensity reflected by the sample and $I_{\mathrm{reference}}$ for the intensity reflected by the CuBe gasket. The data were measured from approx.\ 200\,\cm\ to 17000\,\cm\ (0.03\,eV to 2.11\,eV). Over a small range from app. 1700\,\cm\ to 2700\,\cm, the measured spectra had to be interpolated, as the characteristic multiphonon absorption in diamond is leading to falsified results~\cite{Su.2013}. Also, extrapolations in the low and high energy regime are merged to the spectra. These extrapolations are constructed with the help of literature values from transport measurements and volumetric information, which is necessary to obtain optical functions, like the optical conductivity $\sigma_1$ and the loss function -Im(1/$\epsilon$). This is done by applying the Kramers-Kronig relations through programs by David Tanner~\cite{Tanner.2015}. All data sets are fitted with the Drude-Lorentz model to learn about pressure-dependent behaviour of various electronic excitations with the use of the RefFIT software, as described by Kuzmenko et al.~\cite{Kuzmenko.2005}.

\section{Results}
\subsection{Pressure-dependent optical spectra}
The room-temperature reflectance spectra of MBT as a function of pressure have been measured in small steps up to 8.0\,GPa. In order to establish a good sample-diamond interface, which is necessary for reasonable results, the first spectrum has been taken at 1.1\,GPa. The results are shown in Fig.~\ref{fig.opticaldata}\,(b).
At the lowest pressure, we find a characteristic metallic spectrum: The reflectivity level at the lowest measured frequency is quite high, close to 80\,\%, which is followed by a rather strong decrease corresponding to the plasma edge. The reflectance shows the plasma minimum at app. 1200\,\cm and a weak but steady increase up to the highest measured frequency. With increasing pressure, this profile is maintained. Yet, we observe that the level below $\sim$700\,\cm\ is decreasing with increasing pressure, while above this wavenumber a reflectance increase is found.
The corresponding optical conductivity spectra [see Fig.~\ref{fig.opticaldata}\,(c)] show for all measured pressures a low level
($\sim$1000\,$\Omega^{-1}$\cm) at low frequencies, which is followed by a linear-like increase above approx.\ 2000\,\cm. This quasi-linear increase in $\sigma_1$ has been extrapolated with linear fits, in order to estimate the size of the optical gap, as will be discussed later.
The high-frequency range is dominated by a pronounced absorption band centered at around 10000\,\cm due to interband transitions, which is in good agreement with previous measurements by our group~\cite{Koepf.2020, Koepf.2022, Koepf.2022a} and by Xu et al.~\cite{Xu.2021}. With increasing pressure this absorption band increases in strength and shifts slightly to smaller frequencies.
The inset of Fig.~\ref{fig.opticaldata}\,(b) shows the loss function of MBT as a function of pressure, defined as -Im(1/${\hat{\epsilon}}$) where $\hat{\epsilon}$ is the complex dielectric function.
A clear plasmon peak in the loss function at the lowest pressure indicates the metallic character of the material, and its position corresponds to the value of the screened plasma frequency $\omega_{\mathrm{pl}}^{\mathrm{scr}}$. With increasing pressure the plasmon peak shifts from $\sim$800\,\cm at 1.1~GPa to lower frequencies and broadens considerably, both signalling a weakening of the metallic character of MBT under pressure, as will be discussed below.

Corresponding optical functions of the 26\,\% Sb-doped compound are displayed in Figs.\,\ref{fig.opticaldata}\,(d) and (e) at various pressures up to 10.2 GPa. The overall pressure dependence of the optical response is very similar to that of the undoped material MBT.
However, the metallic character of Mn(Bi$_{0.74}$Sb$_{0.26}$)$_2$Te$_4$ is much more reduced compared to MBT: namely, the low-frequency reflectivity level is low and the plasma edge is less developed with a plasma minimum located at a lower frequency ($\sim$400\,\cm) {see Fig.\,\ref{fig.opticaldata}\,(d)].
Also the Fabry-P\'{e}rot interferences in the frequency range 500 - 1300\,\cm\ signal the much reduced metallic nature, leading to a partial transparency of the studied crystal. The Fabry-P\'{e}rot interferences are reduced under pressure.
At higher frequencies, the level of the reflectivity is rising very weakly, similar to the pure compound [see Fig.~\ref{fig.opticaldata}\,(b)]. The low-frequency optical conductivity $\sigma_1$, depicted in Fig.~\ref{fig.opticaldata}\,(e), has an extremely low level up to frequency 1000\,\cm, confirming the very low spectral weight of free charge carrier excitations. As expected, the material is close to insulating. A small plateau-like feature in $\sigma_1$ at approx.\ 1500\,\cm\ is followed by a steep quasi-linear increase to high values and an absorption band due to interband transitions. The plasmon peak in the loss function of Mn(Bi$_{0.74}$Sb$_{0.26}$)$_2$Te$_4$ [see inset of Fig.~\ref{fig.opticaldata}\,(d)] indicates the pressure-induced weakening of the (already weak) metallic character like for MBT, namely a decrease of its height, broadening, and shift to lower frequencies.

\begin{figure*}[t]
	\includegraphics[width=1\linewidth]{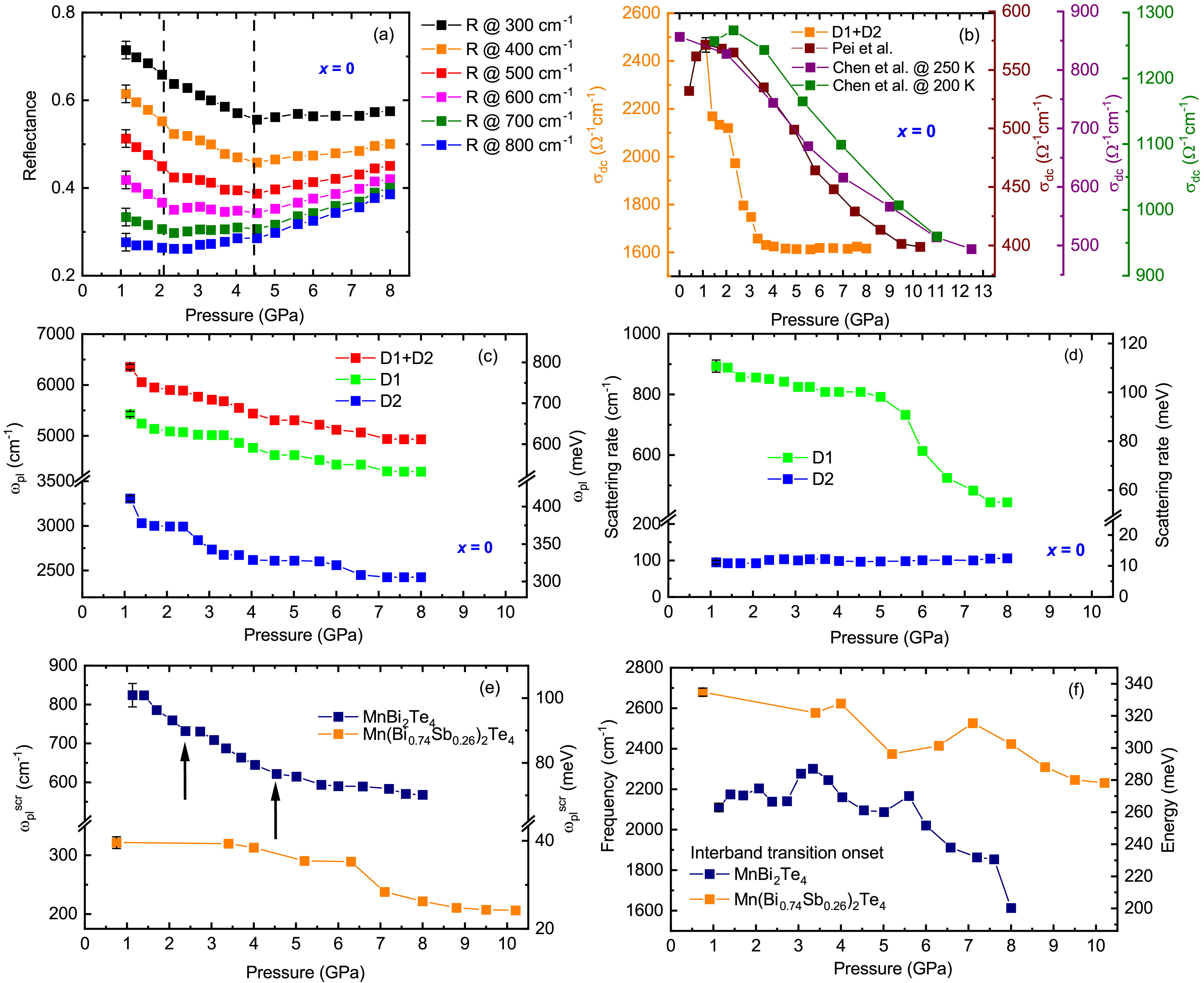}
	\caption{(a) Reflectance values of MBT at selected frequencies as a function of pressure. The dashed, vertical lines highlight the anomalies at $\sim$2 and $\sim$4~GPa. (b) $\sigma_{\mathrm{dc}}$ values of MBT obtained by the Drude-Lorentz fits, which are compared to results from Pei et al.~\cite{Pei.2020} and Chen et al.~\cite{Chen.2019b}. (c) Plasma frequency $\omega_{\mathrm{pl}}$ of the two Drude terms D1 and D2 and the total plasma frequency of the combination D1+D2 (see text for definition) for MBT as a function of pressure. (d) Scattering rate of the Drude terms D1 and D2 for MBT as a function of pressure.
(e) Screened plasma frequency $\omega_{\mathrm{pl}}^{\mathrm{scr}}$ of the pure and the 26\,\% doped sample, as obtained from the plasmon peak position in the loss function. Two anomalies in the pressure dependence of $\omega_{\mathrm{pl}}^{\mathrm{scr}}$ for MBT at $\sim$2 and $\sim$4~GPa are indicated by vertical arrows.
(f) Interband transition onset for MBT and $x$=0.26 doped sample as a function of pressure, determined by the zero-crossing of the linear extrapolations in the $\sigma_1$ spectra, which is an estimate for the size of the optical gap. The error bars in (a) have been estimated within the accuracy of the reflection measurement.}
    \label{fig.parameters}
\end{figure*}

\subsection{Analysis of optical functions and optical parameters as a function of pressure}
For a detailed understanding of the charge carrier dynamics and excitations under pressure, we have performed fits of the measured optical data with the Drude-Lorentz model. Fits of the optical conductivity $\sigma_1$ of MBT at the lowest and highest pressure are depicted in Fig.~\ref{fig.fitting}\,(a) and (b), respectively. As free charge carrier contributions from two electronic bands are expected according to the electronic band structure of MBT [Fig.\ \ref{fig.fitting}\,(c)], and also in order to obtain a reasonable fit of the data, we have implemented two Drude terms besides six Lorentz terms in the measured range. This fit model is in accordance with previous analyzes by our group~\cite{Koepf.2020, Koepf.2022a}.
The ``High-Energy'' oscillator stands for the sum of all oscillators lying outside the measured range. From the comparison of the Drude contributions at lowest and highest measured pressure, we can conclude that the Drude terms are losing spectral weight under pressure, whereas all the Lorentz oscillators are gaining in their spectral weight. The electronic band structure of MBT sketched in Fig.~\ref{fig.fitting}\,(c) is based on calculations by Chen et al.~\cite{Chen.2019a} and has already been discussed in Refs.\ \cite{Koepf.2020, Koepf.2022a}.
The two energetically lowest interband transitions L1 and L2 correspond to transitions between the two conduction bands and to the lowest-energy transitions between the valence and conduction bands, defining the optical gap, respectively.
Especially L1 increases a lot in its oscillator strength when the external pressure is increased, which lifts the level of $\sigma_1$
around 1000~cm$^{-1}$.

For the 26\,\% substituted compound Mn(Bi$_{0.74}$Sb$_{0.26}$)$_2$Te$_4$, the behaviour of the optical conductivity under pressure is partially similar. Comparing the data shown in Fig.~\ref{fig.fitting} at (d) 3.4\,GPa and at (e) 10.2\,GPa the Lorentz oscillators in the high-energy range are gaining in spectral weight. The initial excitation resulting from interband transitions, also referred to as the optical gap, is described by the L2 oscillator, which is located just below the quasi-linear increase. This peak undergoes a slight shift to lower frequencies as well as a narrowing under pressure.
The spectral weight of $\sigma_1$ below $\sim$1000\,\cm\ has a very low level, with a pressure-induced decrease in the Drude spectral weights and an increase in the L1* oscillator strength. L1* describes the electronic transitions between the valence and conduction band as illustrated in Fig.~\ref{fig.fitting}\,(f). As compared to MBT, in Mn(Bi$_{0.74}$Sb$_{0.26}$)$_2$Te$_4$
the energy gap $E_{\mathrm{gap}}$ is reduced and the Fermi level is shifted down in energy and is expected to slightly cut one of the former valence bands~\cite{Chen.2019}. Also, we obvserve an additional oscillator L1*b located at around 200\,\cm, which appears for pressures above 8.8\,GPa. This weak excitation seems to be less screened with increasing pressure, as the spectral weight of the Drude contributions and the overall $\sigma_1$ level are reduced.
To conclude, both samples of this compound family show similar pressure dependences, in particular, a strong reduction of the Drude spectral weights under pressure.

The pressure evolution of several optical parameters is depicted in Fig.~\ref{fig.parameters}, highlighting the most important findings of our study.
In Fig.~\ref{fig.parameters}\,(a) we show the reflectance values for MBT at selected frequencies below 1000 cm$^{-1}$ as a function of pressure. One observes two weak anomalies in the pressure dependence at around 2 and 4.5~GPa, which signal abrupt changes in the electronic properties under pressure.
The pressure-dependent value of the dc conductivity $\sigma_{\mathrm{dc}}$ of MBT as extracted from the Drude-Lorentz \textcolor{red}{fits} is depicted in Fig.~\ref{fig.parameters}\,(b). Obviously, $\sigma_{\mathrm{dc}}$ undergoes a strong decrease from approx.\ 2500\,$\Omega^{-1}$\cm\ at 1~GPa to 1600\,$\Omega^{-1}$\cm\ at 4~GPa followed by a constant behavior. We compare our experimental data to results from Pei et al.~\cite{Pei.2020} measured at room temperature and from Chen et al.~\cite{Chen.2019b} measured on two samples at 250\,K and 200\,K, respectively.
Despite the difference in absolute values, we observe different slopes in the pressure dependence of $\sigma_{\mathrm{dc}}$ compared to our data. Yet in all cases the dc conductivity is decreasing with increasing external pressure. A small pressure-induced increase of the dc conductivity at pressures below 3\,GPa was observed by Pei et al.~\cite{Pei.2020} and for a sample at 200\,K by Chen et al.~\cite{Chen.2019b}, which is followed by a strong decrease above. This initial discrepancy is not visible in our data, where $\sigma_{\mathrm{dc}}$ decreases already from the lowest pressure onwards.

As a measure of the metallic character of a material we can furthermore consider the plasma frequency, which corresponds to the Drude spectral weight, and the scattering rate of the Drude terms, as extracted from the Drude-Lorentz \textcolor{red}{fits}.
In Fig.~\ref{fig.parameters}\,(c) the values of the $\omega_{\mathrm{pl}}$ of the Drude terms D1 ($\omega_{\mathrm{pl,1}}$) and D2 ($\omega_{\mathrm{pl,2}}$) and the total plasma frequency labelled ``D1+D2'' and calculated according to $\omega_{\mathrm{pl}}=\sqrt{\omega_{\mathrm{pl,1}}^2+\omega_{\mathrm{pl,2}}^2}$ are shown as a function of pressure.
The value of $\omega_{\mathrm{pl}}$ (as well as of $\omega_{\mathrm{pl,1}}$ and $\omega_{\mathrm{pl,2}}$) decreases under pressure, clearly indicating the weaking of the metallic character. The scattering rate of the Drude term D1 (see Fig.~\ref{fig.parameters}\,(d)) shows a small decrease up to 5~GPa, followed by a stronger drop at higher pressures. The scattering rate of D2 is approximately pressure independent.

Another characterization of the free charge carrier dynamics is given by the screened plasma frequency $\omega_{\mathrm{pl}}^{\mathrm{scr}}$, which can be obtained from the plasmon peak position in the loss function (see insets of Figs.~\ref{fig.opticaldata}\,(b) and (d)).
$\omega_{\mathrm{pl}}^{\mathrm{scr}}$ is related to the plasma frequency $\omega_{pl}$ according to $\omega_{pl}^{scr}$=$\omega_{pl}$/$\sqrt{\epsilon_{\infty}}$, where $\epsilon_{\infty}$ is the high-frequency value of $\epsilon_1(\omega)$. For the MBT compound $\omega_{\mathrm{pl}}^{\mathrm{scr}}$ is decreasing under pressure, with weak anomalies at $\sim$2 and $\sim$4.5~GPa indicated by arrows in Fig.~\ref{fig.parameters}\,(e). Like for the pressure-dependent reflectance values, these anomalies signal abrupt changes in the electronic structure. For Mn(Bi$_{0.74}$Sb$_{0.26}$)$_2$Te$_4$ the value of $\omega_{\mathrm{pl}}^{\mathrm{scr}}$ is also decreased under pressure.
To conclude, the unscreened and screened plasma frequency are reduced under pressure for both studied materials indicating that they become less metallic for pressures up to approx.\ 10~GPa, in good agreement with published dc transport experiments \cite{Pei.2020, Chen.2019b}. Such a pressure behavior is very unusual, since the overlap of atomic orbitals generally increases under pressure, which leads to an increase in electronic band width and hence improved conductivity.

A further important parameter characterizing the electronic band structure is the size of the optical gap.
In order to extract the optical gap size from our data, we have extrapolated the quasi-linear increase in $\sigma_1$ with a linear extrapolation, as mentioned above and as sketched in Fig.~\ref{fig.opticaldata}\,(c). The zero-crossing of this extrapolation marks the onset of the interband transitions and hence is an estimate of the size of the optical gap. The so-obtained values of the interband transition onset as a function of pressure are plotted in Fig.~\ref{fig.parameters}\,(f) for the pure and the 26\,\% substituted compound.
For MBT we observe an approximately constant value of 2200\,\cm\ up to $\sim$6~GPa followed by a drop to 1600\,\cm\ at 8 GPa.
In the case of Mn(Bi$_{0.74}$Sb$_{0.26}$)$_2$Te$_4$ the interband transition onset decreases from 2700\,\cm\ to 2200\,\cm.
Accordingly, in both materials the optical gap is reduced under pressure.

\section{Discussion}
\subsection{Pressure-induced anomalies in the optical data}
For MBT, we observe weak anomalies at $\sim$2\,GPa and $\sim$4\,GPa in the pressure dependence of the reflectance at selected frequencies [Fig.~\ref{fig.parameters}\,(a)], the dc conductivity $\sigma_{dc}$ [Fig.~\ref{fig.parameters}\,(b)], and of the screened plasma frequency $\omega_{\mathrm{pl}}^{\mathrm{scr}}$ [Fig.~\ref{fig.parameters}\,(e)], suggesting abrupt changes in the electronic band structure. These changes could be interrelated to abrupt pressure-induced changes in the crystal structure reported in the literature.
Chen et al.~\cite{Chen.2019b} observed a sudden drop in the pressure dependence of the lattice parameters $a$ and $c$ at 2~GPa, which was interpreted in terms of a lattice softening. According to experimental and theoretical results in Refs.\ \cite{Pei.2020,Xu.2022}, respectively, the interlayer lattice parameter $c$ is more sensitive regarding pressure than the intralayer lattice parameter $a$ for low pressures up to $\sim$2~GPa. However, this behavior changes for higher pressures, namely above approx.\ 4~GPa, since the interlayer distance is less affected in this pressure range. Accordingly, the lattice parameter ratio $c/a$ initially decreases under pressure and exhibits a minimum between $\sim$2 and $\sim$4~GPa, followed by an almost linear increase above 4~GPa up to approx.\ 14\,GPa \cite{Pei.2020}.
The pressure dependence of the $c/a$ ratio can be related to anomalies in the pressure shift of several structural parameters \cite{Pei.2020}:
For example, various bond distances between the Mn, Bi and Te atoms, like the Mn-Te and Bi-Te bond lengths, are decreased under pressure, with an anomalous behavior between 2 and 4\,GPa. Such a pressure dependence is also revealed by the Te-Te bond length, which is a measure for the interlayer distance, and by the bond angles within the Te-Bi-Te octahedra. Besides, the intensity ratio of the strongest Raman-active modes follow this anomalous pressure dependence. It is important to note that no crystal symmetry change occurs for pressures below 14.6~GPa \cite{Pei.2020}.

Interestingly, in the closely related compounds Bi$_2$Te$_3$, Bi$_2$Se$_3$, and Sb$_2$Te$_3$ a Lifshitz transition \cite{Lifshitz.1960}, i.e., an electronic topological transition with changes in the Fermi surface topology {\it without} a lattice symmetry change, has been reported to occur between 3 and 5\,GPa \cite{Polian.2011,Gomis.2011,Vilaplana.2011a,Vilaplana.2011b}.
The layered, polar semiconductor BiTeI, which turns into a topological insulator under moderate pressure~\cite{Bahramy.2012}, is another example, where an electronic topological transition has been observed between 2 and 3~GPa~\cite{Xi.2013}: The pressure evolution of the $c$/$a$ lattice parameter ratio has a minimum in the absence of lattice symmetry change, and the optical parameters show anomalous behavior, similar to our findings in MBT.
Layered compounds are generally prone to pressure-induced electronic topological transitions \cite{Zhu.2012,Yang.2017,Bassanezi.2018} as also evidenced in BiTeBr \cite{Ohmura.2017}, 1T-TiTe$_2$ \cite{Rajaji.2018}, and ZrSiTe \cite{Ebad-Allah.2019,Krottenmuller.2020}.
A topological phase transition has indeed been theoretically predicted for MBT under hydrostatic tensile strain \cite{Guo.2021} and suggested to occur in pressurized MBT \cite{Pei.2020} and in the doping series (Mn$_{1-x}$Pb$_x$)Bi$_2$Te$_4$ \cite{Qian.2022}.
The closely related material MnSb$_2$Te$_4$ undergoes a pressure-induced topological phase transition, when the interlayer distance is decreased by a critical percentage \cite{Zhou.2020}. Interlayer interaction therefore seems to be the driving mechanism for this phase transition. Considering the anomalous behavior of the $c$/$a$ ratio in MBT under pressure as described above, a similar scenario might hold here as well.
The occurrence of a Lifshitz transition in pressurized MBT due to an enhanced interlayer interaction thus seems very likely. In the case of Mn(Bi$_{0.74}$Sb$_{0.26}$)$_2$Te$_4$ we cannot draw a conclusion regarding anomalies in the low-energy optical response, since this compound is less metallic (in fact, close to insulating) and therefore the pressure-induced changes are much less developed.

\subsection{Optical gap evolution}
The energy gap evolution might also be linked to these pressure-induced structural anomalies. Based on theoretical calculations, Xu et al.~\cite{Xu.2022} found a slight increase of the energy gap in MBT up to approx.\ 2\,GPa, followed by an almost linear decrease and a gap closing slightly above 15\,GPa. We cannot directly extract the energy gap from our optical data, since it is smaller than the optical gap, i.e., the onset of interband transitions [see energy scheme in Fig.~\ref{fig.fitting}\,(c)]. The size of the optical gap can only serve as an upper bound for the energy gap size. According to our optical results presented in Fig.~\ref{fig.parameters}\,(e), the optical gap is decreasing with increasing pressure for both studied compounds.

\subsection{Charge carrier dynamics under pressure}
The weakening of the metallic character of MBT under pressure is puzzling and possible reasons for it have been discussed in the literature.
Pei et al.~\cite{Pei.2020} ascribed the pressure-induced increase in resistivity to the competition between the localization of the surface electrons and the delocalization of the bulk electrons, where the first process is predominant in the pressure range between 3 and approx.\ 15\,GPa resulting in a resistivity increase.
Such a resistivity increase, i.e., decrease in metallic character is confirmed by our optical results. However, our optical data do not confirm this scenario: Since infrared spectroscopy is generally bulk sensitive with only minor contributions from surface electrons, the observed pressure-induced decrease in conductivity is a bulk property according to our optical results.

Based on Hall effect measurements, Chen et al.~\cite{Chen.2019b} related the pressure-induced increase in resistivity to a decrease in electron mobility $\mu_{\mathrm{e}}$, despite the observed increase in charge carrier density $n_{\mathrm{e}}$ under pressure.
Two scenarios were discussed as possible explanations \cite{Chen.2019b}. In the first scenario, the pressure-induced suppression of long-range antiferromagnetic order in MBT causes an enhancement of magnetic fluctuations, which increases the electron scattering.
The electron scattering rate $\gamma$ is related to the mobility $\mu_{\mathrm{e}}$ according to
\begin{equation}\label{equation-mobility}
    \mu_{\mathrm{e}}= \frac{e}{m^*_{\mathrm{e}}\gamma} \quad ,
\end{equation}
where $m^*_{\mathrm{e}}$ denotes the effective electron mass~\cite{Yu.2010, Wooten.1972}.
Accordingly, an enhanced scattering rate would decrease $\mu_{\mathrm{e}}$ and hence lead to localization of charge carriers with an enhancement of the resistivity. Since the carrier scattering determines the width of the Drude terms in our optical conductivity data, we can test this scenario. As shown in Fig.\ \ref{fig.parameters}\,(d) the scattering rate of the dominant Drude term D1 is decreasing with increasing pressure, with a stronger drop above $\sim$5~GPa.
Interestingly, above this pressure a suppression of the antiferromagnetic ordering occurs \cite{Pei.2020,Chen.2019b}.
For the weaker Drude contribution D2 the scattering rate is pressure independent.
Therefore, our optical results do not support an enhancement of the scattering rate under pressure, as suggested in Ref.~\cite{Chen.2019b}.

According to Eqn.~\ref{equation-mobility} the mobility of the carriers can also be affected by their effective mass: A pressure-induced increase in $m^*_{\mathrm{e}}$ would cause a decrease in $\mu_{\mathrm{e}}$ consistent with the Hall effect experiments. In fact, an increase in $m^*_{\mathrm{e}}$ would also explain the pressure-induced decrease in the plasma frequency $\omega_{pl}$ [see Fig.\ \ref{fig.parameters}\,(c)], since $\omega_{pl}$ depends on $m^*_{\mathrm{e}}$ according to \cite{Fox.2001}
\begin{equation}
    \omega_{\mathrm{pl}} = \sqrt{\frac{n_{\mathrm{e}} e^2}{\varepsilon_0 m^*_{\mathrm{e}}}} \quad .
\end{equation}
Thus, an increase in $m^*_{\mathrm{e}}$ under pressure could explain both the observed decrease in carrier mobility and decrease in plasma frequency, despite the increase in electron density $n_{\mathrm{e}}$~\cite{Chen.2019b}.
A pressure-induced enhancement of the effective electron mass would be consistent with the second scenario proposed in Ref.\ \cite{Chen.2019b}: In analogy to CaMn$_2$Bi$_2$ \cite{Gibson.2015, Lane.2023, Piva.2019}, pressure application could induce a partial delocalization of Mn $3d$ electrons and subsequent hybridization with Bi $6p$ and/or Te $5p$ conduction electrons, causing a localization of the electrons and the opening of a hybridization gap \cite{Chen.2019b}. The scenario of a pressure-induced localization of charges due to hybridization could explain the results from Hall effect measurements as well as our optical data for both studied materials.

\section{Conclusion}
In this work, we have studied the charge dynamics of the topological insulators MnBi$_2$Te$_4$ and Mn(Bi$_{0.74}$Sb$_{0.26}$)$_2$Te$_4$ at high pressures by determining the optical response functions. In good agreement with other pressure studies on MnBi$_2$Te$_4$~\cite{Pei.2020, Xu.2022, Chen.2019b, Guo.2021, Shao.2021}, we have detected a pressure-induced decrease in the metallic strength for both materials based on the low-energy optical response. This decrease in conductivity is possibly due to an enhanced effective mass resulting from the creation of a hybridization gap. Furthermore, several optical parameters show weak anomalies
in their pressure dependence at $\sim$2 and $\sim$4~GPa,  which might originate from a topological electronic transition. In analogy to the closely related material MnSb$_2$Te$_4$, we suggest the enhancement of interlayer interaction as the driving mechanism for this phase transition. Besides, we observe a pressure-induced decrease in the optical gap size for both studied materials, indirectly confirming the expected reduction of the energy gap under pressure.


\section{Acknowledgments}
The authors acknowledge the fruitful discussions with and the technical support by Jihaan Ebad-Allah.
C.\ A.\ K.\ acknowledges financial support from the Deutsche
Forschungsgemeinschaft (DFG), Germany, through Grant
No. KU 1432/15-1.  Z.Q.M. and S.H.L. acknowledges the support of the US NSF through the Penn State 2D Crystal Consortium-Materials Innovation Platform (2DCC-MIP) under NSF Cooperative Agreement DMR-2039351.

\end{document}